\documentclass[citeautoscript,floatfix,aps,prl,twocolumn,
      superscriptaddress]{revtex4-1}

\usepackage{amsmath}
\usepackage{graphicx}
\usepackage{epstopdf}
\usepackage{natbib}
\usepackage{array}
\usepackage[usenames,dvipsnames]{color}
\usepackage{dcolumn}
\usepackage{bm}
\usepackage{soul}  
\usepackage{changes}

\def\nk{{n\textbf{k}}}

\def\mk{{m\textbf{k}}}
\def\unk{u_{n\textbf{k}}}
\def\umk{u_{m\textbf{k}}}

\def\bfk{{\textbf{k}}}
\def\sns{SnS$_2$}

\begin{document}
\title{Nonadiabatic Born effective charges in metals and the Drude weight}
\author{Cyrus E. Dreyer}
\affiliation{Department of Physics and Astronomy, Stony Brook University, Stony Brook, New York, 11794-3800,
  USA}
\affiliation{Center for Computational Quantum Physics, Flatiron Institute, 162 5th Avenue, New York, New York 10010,
USA}
\author{Sinisa Coh}
\affiliation{Materials Science and Mechanical Engineering, University of California Riverside, CA 92521, USA}
\author{Massimiliano Stengel}
\affiliation{Institut de Ci\`{e}ncia de Materials de Barcelona (ICMAB-CSIC), Campus UAB, 08193 Bellaterra, Spain}
\affiliation{ICREA-Instituci\'{o} Catalana de Recerca i Estudis Avan\c{c}ats, 08010 Barcelona, Spain}
\date{\today}

\begin{abstract}
In insulators, Born effective charges describe the electrical polarization induced by the displacement of individual atomic sublattices.
Such a physical property is at first sight irrelevant for metals and doped semiconductors, where the macroscopic polarization is ill-defined.
Here we show that, in clean conductors, going beyond the adiabatic approximation results in nonadiabatic Born effective charges that are well defined in the low-frequency limit. 
In addition, we find that the sublattice sum of the nonadiabatic Born effective charges does not vanish as it does in the insulating case , but instead is proportional to the Drude weight.
We demonstrate these formal results with density functional perturbation theory calculations of Al, and electron-doped \sns{} and SrTiO$_3$. 
\end{abstract}

\maketitle

Born effective charges (BECs) are defined as the polarization induced by the displacement of an atomic sublattice \cite{Born1954,Lyddane1941,Pick1970}. 
They are key quantities for the microscopic understanding of a wide variety of phenomena in insulators, 
including the interplay between electric fields and long-wavelength optical phonons \cite{Born1954,GonzeLee1997,Vogl1976,Sjakste2015,Verdi2015}, lattice contribution to dielectric screening \cite{Born1954,GonzeLee1997} and electromechanical coupling \cite{Martin1972,Hong2013,Stengel2013}, and polar properties of ferroelectrics \cite{Resta1993,Zhong1994,Ghosez1998}.
BECs (also known as dynamical charges) were placed on a firm conceptual footing in the seminal work of Pick, Cohen and Martin (PCM)~\cite{Pick1970}, which formally established the acoustic sum rule (ASR), that a rigid translation of the crystal lattice does not produce either forces on the individual atoms, or a net electrical polarization. This is a consequence of translational invariance and charge neutrality, and requires that the sublattice sum of the BECs vanish.

The macroscopic polarization and its parametric derivatives are only well defined in a gapped system~\cite{Resta1994}, which at first sight rules out the calculation of BECs in metals. 
In addition, a hypothetical definition of the BECs as
the dipolar moment of the first-order charge induced by an atomic displacement would also be problematic, since the free carriers screen any long-range electrostatic perturbation.
Yet, the concept of BEC is routinely used to study doped (and hence metallic) semiconductors, where the Fr\"ohlich divergence~\cite{Verdi2015,Sjakste2015} in electron-phonon matrix elements plays a central role in carrier scattering; 
the BECs for doped semiconductors are calculated in the undoped limit,~\cite{Verdi2015,Sjakste2015} but the validity of such an approximation remains to be seen.
Also, the recent surge of interest in ``ferroelectric metals'' \cite{Anderson1965,Shi2013} has provided additional motivation to understand BECs in metallic systems.

As we shall see shortly, one can circumvent the issues with defining and calculating BECs in metallic systems by 
relaxing
the adiabatic (or Born-Oppenheimer \cite{Born1927}) approximation,
a key assumption in the PCM derivations.
The adiabatic approximation is usually justified in insulators, where the electronic gap is much larger than typical optical phonon frequencies ($<200$ meV for inorganic materials \cite{Petretto2018}), but
generally breaks down in a metallic or doped semiconducting systems.
Manifestation of nonadiabaticity in lattice dynamics are well known, e.g., plasmon-phonon coupling~\cite{Varga1965,Olson1969} and renormalization of the phonon frequencies~\cite{Calandra2010} .
In particular, there exists a region near the zone center where the phase velocity of an optical phonon is large compared to the Fermi velocity ~\cite{Engelsberg1963,Maksimov1996,Saitta2008}. In such a regime, the nonadiabatic dynamical matrix may significantly differ from its adiabatic counterpart. Also, free carriers are unable to screen the long-range electric fields~\cite{Pisana2007}, and signatures of the BECs can be detected experimentally as 
resonances in the reflectivity spectra~\cite{Bistoni2019,Binci2020} or shifts in the plasmon frequency~\cite{Varga1965}.

In this work, we show that nonadiabatic BECs (naBECs) are well-defined real quantities (as long as time-reversal symmetry is preserved) in the low-frequency regime relevant for lattice dynamics.
Remarkably, we find that their sublattice sum in general does not vanish but tends
to the \emph{Drude weight} (DW), which gives the density of free electrons available for conduction.
This generalized sum rule implies that the naBECs never vanish in metals or doped semiconductors (not even in elemental quasi-free-electron crystals).
This result, in addition to clarifying the origin of the ASR breakdown that was pointed out in Ref.~\onlinecite{Binci2020}, provides a novel physical interpretation of the DW in terms of lattice dynamics (as opposed to electrical transport).
Indeed, the DW emerges as the mechanical inertia of the conduction electrons to an 
acceleration of the underlying lattice potential, which cannot drag the carriers along 
if subjected to a sufficiently rapid oscillation.
We demonstrate our formal results with a density-functional perturbation theory (DFPT) \cite{Baroni2001,Giustino2017} based methodology for calculating the naBECs and DW for a prototypical metal (Al) and two doped semiconductors (\sns{} and SrTiO$_3$).
In the latter, we find that naBECs 
can deviate significantly from their values in the undoped material.

We shall frame our discussion in terms of generalized susceptibility functions,
$\chi_{\lambda_1\lambda_2}(\omega,\textbf{q})$, describing the $\lambda_1$-response to the perturbation 
$\lambda_2$, modulated by a wavevector \textbf{q} and at a frequency $\omega$. 
We focus henceforth on the cases where $\lambda_{1,2}$ are either an atomic displacement of sublattice $\kappa$ in direction $\alpha$ ($\tau_{\kappa \alpha}$) or a component of the vector potential ($A_\alpha$).
The purely electromagnetic case is well known: 
$\chi_{A_\alpha A_{\beta}}(\omega,\textbf{q})$
is the current--current response~\cite{Adler1962},
whose long-wavelength limit relates to the macroscopic optical conductivity via~\cite{Allen}
\begin{equation}
 \label{sigmaw}
    \begin{split}
     \sigma_{\alpha\beta}(\omega) &=
     \frac{i}{\omega}
     \lim_{{\bf q} \rightarrow 0}
\chi_{A_\alpha A_{ \beta}}(\omega,{\bf q}). 
        \end{split}
\end{equation}
(As customary, we shall assume that the frequency has a small imaginary part,
$i\eta$, to ensure causality~\cite{Resta2018}.) 
In close analogy with Eq.~(\ref{sigmaw}), we shall define naBECs as \cite{Bistoni2019,Binci2020}
\begin{equation}
 \label{zw}
    \begin{split}
     Z^{(\alpha)}_{\kappa\beta}(\omega) &=
     -\frac{i}{\omega}
     \lim_{{\bf q} \rightarrow 0}
\chi_{A_\alpha \tau_{\kappa \beta}}
 (\omega,{\bf q}),
        \end{split}
\end{equation}
which has the physical meaning of a current response to the atomic \emph{velocity}~\footnote{The negative sign compared to Eq.~(\ref{sigmaw}) reflects the fact that the time derivative of the vector
potential is \emph{minus} the electric field, ${\bf E} = -\partial_t {\bf A}$}.

For most materials, the interesting physics is
contained in the small-$\omega$ behavior of $Z^{(\alpha)}_{\kappa\beta}(\omega)$,
i.e., at the frequencies that are relevant for lattice dynamics.
Within such a regime, the conductivity [Eq.~(\ref{sigmaw})] vanishes
in topologically trivial insulators~\cite{Thouless1982,Resta2018} 
and
Eq.~(\ref{zw}) reduces to the standard linear-response 
formula~\cite{Baroni2001,Gonze1997,GonzeLee1997} for 
the \emph{adiabatic} BEC. 
Note that, in the insulating case, the result is also independent 
of the order of the long-wavelength and small-frequency limits.

In metals, Eq.~(\ref{sigmaw}) diverges as $\omega \rightarrow 0$, 
resulting in the Drude peak in the longitudinal conductivity~\cite{Allen}.
Since Eq.~(\ref{zw}) is formally similar to Eq.~(\ref{sigmaw}),
it is reasonable to wonder whether 
$Z^{(\alpha)}_{\kappa\beta}(\omega \rightarrow 0)$ diverges as well,
or tends to a finite constant. 
The divergence in $\sigma_{\alpha\beta}(\omega \rightarrow 0)$ is 
rooted in the fact that $\chi_{A_\alpha A_\beta}$ 
tends to a finite nonzero limit in all metals, 
\begin{equation}
\label{eq:DW}
\lim_{\omega \rightarrow 0} \chi_{A_\alpha A_\beta}(\omega) = \frac{D_{\alpha \beta}}{\pi},
\end{equation}
where $D_{\alpha\beta}$ is the DW tensor~\cite{Resta2018}.
The situation regarding the mixed response to a vector potential and
an atomic displacement clearly differs, as it requires time-reversal (TR) symmetry
to be broken in order to be nonzero. 
(${\bf A}$ is odd with respect to both TR and space inversion,
while $\bm{\tau}_\kappa$ is TR-even.)
This simple argument guarantees that in TR-symmetric metals, 
where a steady current in response to a static atomic displacement is forbidden,
$\chi_{A_\alpha \tau_{\kappa\beta}}(\omega \rightarrow 0)$ vanishes
and therefore
$Z^{(\alpha)}_{\kappa\beta}(\omega \rightarrow 0)$ is well defined.
Note that, unlike the insulating case, here the 
order of the 
${\bf q}\rightarrow 0$ and $\omega \rightarrow 0$ limits do not commute; 
our prescription of taking the ${\bf q}\rightarrow 0$
first
is relevant for optical phonons,
which retain a finite frequency in the long-wavelength limit.

Acoustic phonons are, in principle, not concerned by 
our arguments, since their frequency vanishes linearly 
with momentum in a neighborhood of the zone center.
Yet, in the following we shall consider hypothetical long-wavelength
acoustic phonons whose limiting ${\bf q} \rightarrow 0$ frequency is 
artificially assumed to remain finite.
This assumption, while physically unrealistic, will serve as an 
intermediate step towards the derivation of a sum 
rule for the nonadiabatic BECs, which shall be our next goal.
The rationale behind this procedure is that an acoustic phonon at
\textbf{q}=0 reduces to a rigid translation of
the whole crystal lattice.
In the linear regime, a translation can be regarded as the 
sum of the displacements of all individual atomic sublattices;
thus, by looking at the current density induced by a ``zone-center
acoustic phonon'' that is externally modulated in time at a frequency 
$\omega$, one can directly infer the sublattice sum of the naBECs [Eq.~(\ref{zw})].

The advantage of such an approach is that the acoustic phonon perturbation enjoys a remarkably compact expression when rewritten, via a coordinate transformation,
in the curvilinear frame that is co-moving with the atoms~\cite{Stengel2013natcom,Stengel2018,Schiaffino2019}. 
Thereby, an acoustic phonon can be recast as a ``static'' strain contribution that only depends on the metric of the deformation, plus
a ``dynamical'' effective vector potential that
results from the inertia of the electrons 
upon local displacements of the coordinate frame~\cite{Stengel2018}. In a Kohn-Sham \cite{Kohn1965} density-functional context (see supplemental material (SM)~\cite{SM} Sec.~S1), we can then write
the sublattice sum of the first-order phonon ($\tau_{\kappa\beta}$) Hamiltonians as~\cite{Stengel2018}
\begin{equation}
\label{metric}
\sum_\kappa \hat{\mathcal{H}}^{\tau_{\kappa\beta}}(\omega,{\bf q}) 
  = \hat{\mathcal{H}}^{(\beta)}(\omega,{\bf q})  + i \omega \hat{\mathcal{H}}^{A_\beta}(\omega,{\bf q}),
\end{equation}
where $\hat{\mathcal{H}}^{(\beta)}(\omega,{\bf q})$ is the ``metric wave'' Hamiltonian of Refs.~\citenum{Stengel2018} and \citenum{Schiaffino2019}, and $\hat{\mathcal{H}}^{A_\beta}(\omega,{\bf q})$ is the additional effective vector potential perturbation.

As a result of translational invariance, $\hat{\mathcal{H}}^{(\beta)}(\omega,{\bf q})$
identically vanishes at the zone center~\cite{Schiaffino2019}, and thus the rigid displacement of the lattice reduces to the vector-potential perturbation term at $\textbf{q}=0$. 
By combining this observation with Eqs.~(\ref{sigmaw}) and (\ref{zw}), we
readily arrive at 
\begin{equation}
 \label{sumrulew}
    \begin{split}
  \frac{1}{\Omega} \sum_\kappa Z^{(\alpha)}_{\kappa\beta}(\omega) &= i\omega \sigma_{\alpha \beta}(\omega),
        \end{split}
\end{equation}
relating the sum of the naBECs to the optical 
conductivity ($\Omega$ is the cell volume).
From Eq.~(\ref{eq:DW}) we find that the small-frequency limit of Eq.~(\ref{sumrulew}) becomes,
\begin{equation}
  \label{sumruleD}
\frac{1}{\Omega} \sum_{\kappa} Z^*_{\alpha,\kappa\beta} =\frac{D_{\alpha\beta}}{\pi},
\end{equation}
where $Z^*_{\alpha,\kappa\beta}\equiv Z^{(\alpha)}_{\kappa\beta}(\omega \rightarrow 0)$ is the zero-frequency limit of the naBEC.
We call Eq.~(\ref{sumruleD}) the \emph{naBEC sum rule for metals} and, along with Eq.~(\ref{sumrulew}), is the central result of this work.
Of course, in an insulator $D_{\alpha\beta}$ 
vanishes, and we recover the well-known version of the acoustic sum rule \cite{Pick1970}, $\sum_{\kappa}Z^*_{\alpha,\kappa\beta}=0$, 
The physical interpretation of Eq.~(\ref{sumruleD}) is 
that the DW quantifies
the portion of the electron charge that is ``free,'' i.e., 
not bound to the underlying atomic lattice. 
When the crystal is
rigidly translated, going beyond the adiabatic assumption means
that exactly this charge is ``left behind,'' resulting in a positive
current due to the displacement of the uncompensated nuclear
charges. This provides an alternative physical interpretation of the
DW, which is based on lattice dynamics rather than
electronic transport.

The above results are relevant for the clean limit of a metal or semiconductor, 
i.e., we assume an optical phonon of frequency $\omega$ that is much larger that
the inverse carrier lifetime, $1/\tau$
\cite{Calandra2010,Saitta2008}.
Thus, the limit $Z^{(\alpha)}_{\kappa\beta}(\omega \rightarrow 0)$ in Eq.~(\ref{sumruleD}) should not be taken literally: the
result only holds for a window of frequencies where $\omega\gg 1/\tau$,
and both 
parameters are far smaller than interband resonances.
Whenever the clean condition breaks down, we expect the naBECs to vanish as
the carriers have enough time to relax to the instantaneous electronic
ground state along the phonon displacement coordinate. This is consistent
with the arguments of Refs.~\citenum{Maksimov1996}, \citenum{Calandra2010}, and \citenum{Saitta2008}, made for nonadiabatic corrections to the phonon frequencies.
Interestingly, Eq.~(\ref{sumrulew}) appears to be qualitatively correct even
in presence of dissipation -- the DC limit of the conductivity is finite,
which implies that the ASR is recovered for $\omega \rightarrow 0$.
%
Deriving the nonadiabatic current in the co-moving reference frame 
was key to achieving the qualitatively correct results, as it reflects the
intuitive idea that scattering events tend to equilibrate the average momentum
of the carriers with respect to the underlying crystal lattice, rather than
the laboratory frame~\cite{vanLeeuwen2004}.

We will now demonstrate the sum rule in Eq.~(\ref{sumruleD}) with DFPT calculations performed on three systems,  the face-centered cubic (FCC) phase of Al, electron-doped bulk ($P\overline{3}m1$) SnS$_2$ and electron-doped cubic ($Pm\overline{3}m$) SrTiO$_3$ (STO). The details of the implementation, computational parameters, and numerical convergence for these calculations are given in the SM \cite{SM} Secs.~S1 D and S2. The doping in SnS$_2$ and STO is applied under the ``rigid band'' approximation, i.e., all matrix elements and structural parameters are taken from an undoped calculation, and just the occupation factors 
are changed to reflect the added electrons; the excess positive charge to compensate the electrons is added to the ionic BEC of a given sublattice, corresponding to chemical doping on that sublattice. In SM \cite{SM} Sec.~S2 C, we compare these results to explicit doping via the virtual crystal approximation, demonstrating that the naBEC sum rule is satisfied regardless of the approximation.

We calculate the naBECs in the  $\omega\rightarrow0$ limit as (see, Sec.~S1~\cite{SM} and Refs.~\cite{Bistoni2019,Binci2020})
\begin{widetext}
\begin{equation}
  \label{eq:NABEC}
  \begin{split}
  Z^*_{\alpha,\kappa\beta}&=Z^{\text{ion}}_{\kappa}\delta_{\alpha\beta}
  - \text{Im} \lim_{\eta \rightarrow 0^+}
  \int [d^3 k]
    \sum_{n\neq m}
 \frac{f_{\nk}-f_{\mk}}
    {(\epsilon_\nk-\epsilon_\mk)
     (\epsilon_\nk-\epsilon_\mk + i \eta)}
    \langle\unk | \hat{H}_{\textbf{k}}^{A_\alpha}
    \vert\umk\rangle \langle\umk\vert
    \hat{\mathcal{H}}^{\tau_{\kappa \beta} 
    }_{\textbf{k}}\vert\unk\rangle,
\end{split}
\end{equation}
\end{widetext}
where $Z^{\text{ion}}_{\kappa}$ is the ionic (i.e., pseudopotential) charge, ${\epsilon}_{\nk}$ and ${f}_{\nk}=f({\epsilon}_{\nk})$ are the
energy and Fermi occupation factor of band 
$n$ and $k$-point $\bfk$,
$u_{n\bfk}$ is the cell-periodic part of the corresponding Bloch function, $\hat{H}_{\textbf{k}}^{A_\alpha}$ is the $\bfk$-derivative of the Hamiltonian, and $\hat{\mathcal{H}}^{\tau_{\kappa \beta}}_{\textbf{k}}$ is the screened (i.e., including self-consistent fields) first-order phonon Hamiltonian.
The DW, in turn, is calculated via \cite{Resta2018}
\begin{equation}
  \begin{split}
    \label{eq:drude}
    \frac{\Omega}{\pi}D_{\alpha\beta}  &=-\int[d^3k]
   \sum_n
    \frac{\partial {f}(\epsilon_{\nk})}
    {\partial \epsilon}
    v_{\nk}^{(\alpha)}   v_{\nk}^{(\beta)},
    \end{split}
\end{equation}
where $v_{\nk}^{(\alpha)} = \langle\unk\vert\hat{H}^{A_\alpha}_{\textbf{k}}\vert\unk\rangle$ is the $\alpha$ component of the band velocity. 
Note that, because of a technical subtlety related to the use of nonlocal pseudopotentials (nlPSP) in the calculations, the sum rule of Eq.~(\ref{sumruleD}) is exactly satisfied only if the second band velocity in Eq.~(\ref{eq:drude}) is replaced with the canonical momentum, $\widetilde{v}_{\nk}^{(\beta)} = \langle\unk\vert\hat{p}_{\beta\textbf{k}}
\vert\unk\rangle$. 
[We shall refer to this revised version of Eq.~(\ref{eq:drude}) as ``modified'' DW, $\widetilde{D}_{\alpha \beta}$ in the following.]
We ascribe this outcome to the well-known ambiguities that arise when combining nlPSPs with electromagnetic fields~\cite{ICL2001,Pickard2003,Dreyer2018};
in practice, we find that its quantitative impact 
is small.

\begin{figure}
\includegraphics[width=\columnwidth]{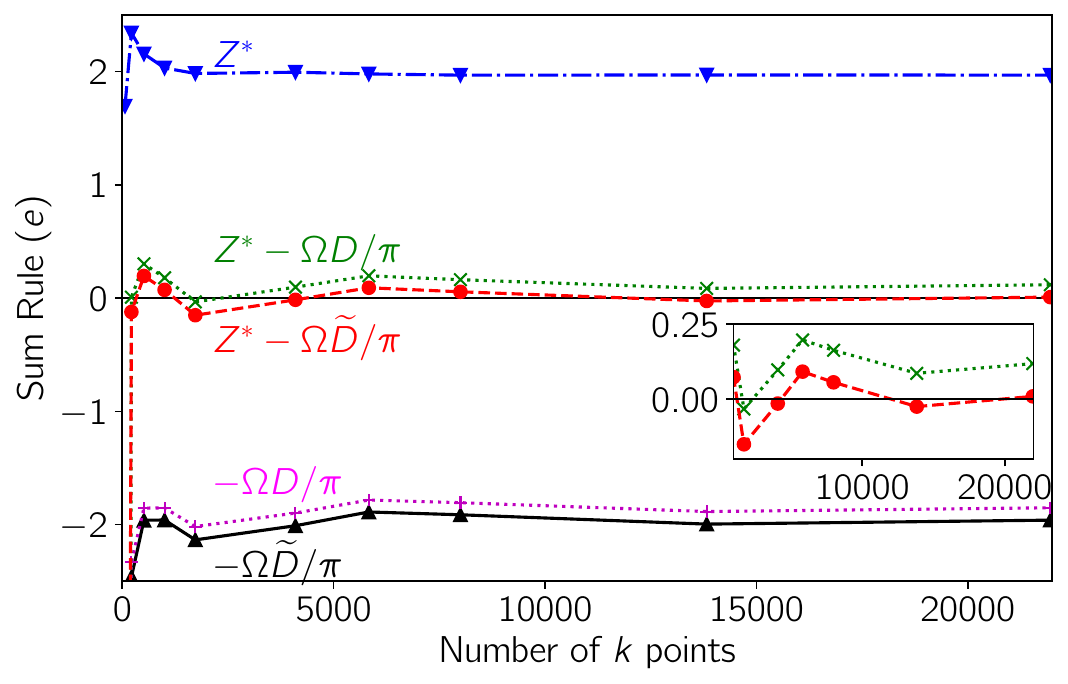}
\caption{\label{Alsumrule} Demonstration of nonadiabatic Born effective charge (naBEC) sum rule for Al versus $k$ points in the Brillouin zone of the conventional 4-atom cell. Blue dot-dashed curve is the naBEC, the black solid (pink dotted) curve is minus the cell volume multiplied by the modified (normal) Drude Weight, and red dashed (green dotted) curve is the sum of the two.}
\end{figure}

In Fig.~\ref{Alsumrule} we plot the naBEC $Z^*$,
standard DW $D$, and modified DW $\widetilde{D}$,
for FCC Al versus the number of $k$-points used to sample the Brillouin zone of the conventional cubic cell~\footnote{Note that the cubic symmetry means
$Z^*_{x,\kappa x}=Z^*_{y,\kappa y}=Z^*_{z,\kappa z}\equiv Z^*$ and, e.g., $D_{xx}=D_{yy}=D_{zz}\equiv D$.}. 
Since FCC Al has no optical modes, the nonadiabatic regime is hardly relevant here from a physical point of view; yet, its computational simplicity allows for a numerical test of our arguments.
Remarkably, the naBEC of Al does not 
vanish in spite of it being an elemental metal,
and converges to a value of around 2$e$.
We see from the red dashed curve that the sum rule in Eq.~(\ref{sumruleD}) is accurately satisfied for a large enough $k$ mesh.
We also see that $\widetilde{D}$ differs slightly from $D$, 
and this results in a violation of the naBEC sum rule that is modest ($\sim0.1$ $e$) but clearly discerned for the numerical accuracy of our calculations. 
We find a similar qualitative behavior for all materials in this study (see, e.g., SM~\cite{SM} Figs.~S9-S11); thus, we shall exclusively focus on $\widetilde{D}_{\alpha\beta}$ henceforth.

\begin{figure}
\includegraphics[width=\columnwidth]{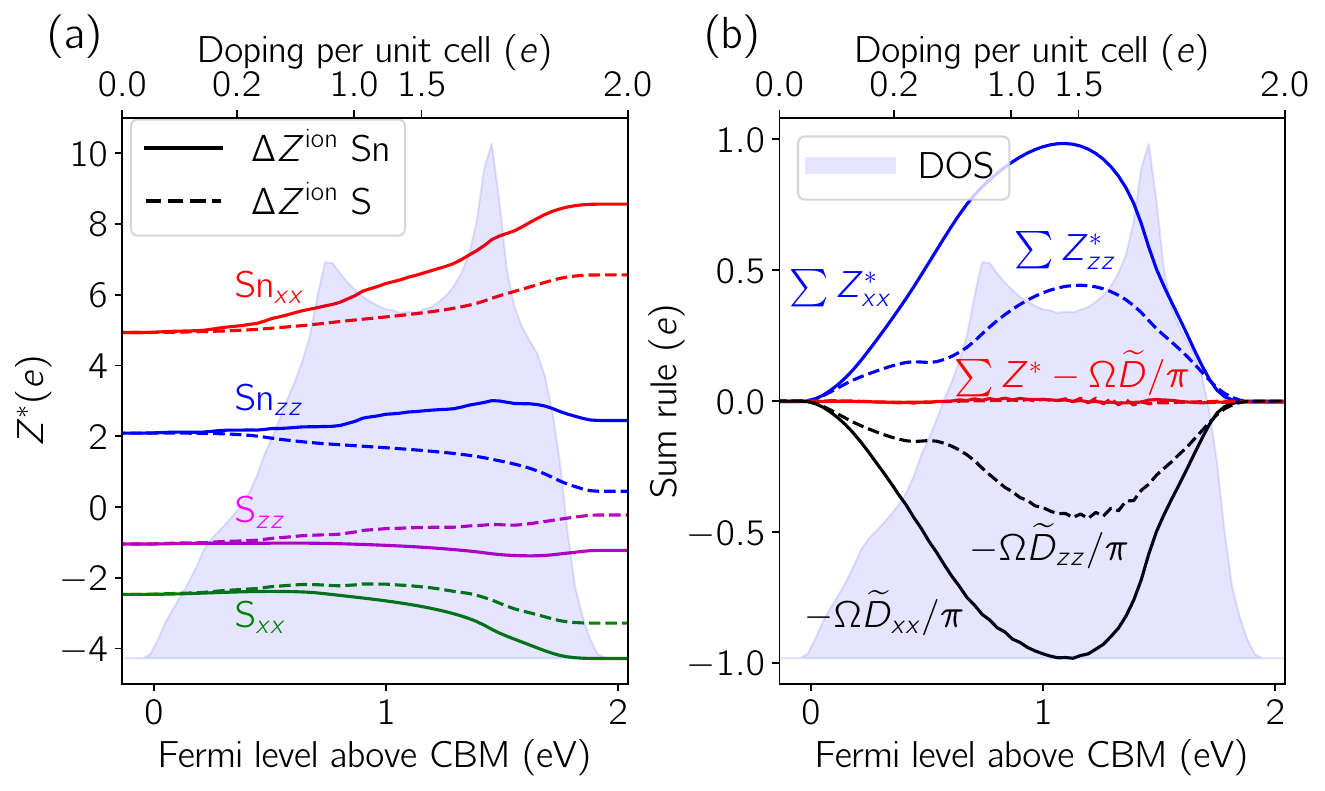}
\caption{\label{SnS2_phi} (a) Nonadiabatic Born effective charges (naBEC) versus doping in the in-plane $x$ direction and out-of-plane $z$ direction. Solid (dashed) lines are for the excess ionic charge associated with the Sn (S) sublattice. (b) Demonstration of the naBEC sum rule, where the sublattice sum of the BECs equals the Drude weight. Blue shaded region is the DOS of the conduction band calculated from the Wannier interpolation.
}
\end{figure}

We now consider bulk \sns{}, a
layered insulator that has shown considerable promise for use in a wide range of device applications \cite{Seo2008,Luo2012,Song2013,Song2013_FE,Yu2014,Chu2018}
We dope \sns{} with electrons into the isolated, lowest-energy Sn($5s$)-S($3p$) conduction band (see Fig.~S2(b) in the SM~\cite{SM}).
In order to accurately describe the  
Fermi surface,  
we perform the Brillouin-zone integrals of 
Eqs.~(\ref{eq:NABEC}) and (\ref{eq:drude}) via
Wannier interpolation to a dense $k$ mesh (see SM~\cite{SM} Secs.~S2 B). 

In Fig.~\ref{SnS2_phi}(a), we show the naBECs versus doping for the Sn and S sublattices, in the in-plane $x$ and out-of-plane $z$ directions. 
The dotted curves in Fig.~\ref{SnS2_phi}(a) correspond to the excess ionic charge placed on the S sublattices, and the solid curves to the charge placed on the Sn sublattice.
In Fig.~\ref{SnS2_phi}(b) we see that the violation of the (sublattice-summed) BEC charge neutrality condition increases as we dope to roughly half of the conduction-band width, and then decreases back to zero when the conduction band is filled. The DW exhibits the same behavior, and we see from the red curves around zero, that the naBEC sum rule [Eq.~(\ref{sumruleD})] is accurately satisfied.

Finally we calculate the naBECs and DW for the electron-doped cubic phase of STO, which has attracted significant attention due to its superconducting properties \cite{Schooley1965,Huang2017,Collignon2019}.
%
To illustrate the behavior of the naBECs, we present doping across the entire Ti $t_{2g}$ conduction band manifold (see Fig.~S2(a)~\cite{SM}), even though it is not expected to be experimentally achievable. (Also, we should note that significant changes to the electronic and atomic structure are expected if such calculations at high doping are performed with atomic relaxations \cite{Bjaalie2014}.)

\begin{figure}
\includegraphics[width=\columnwidth]{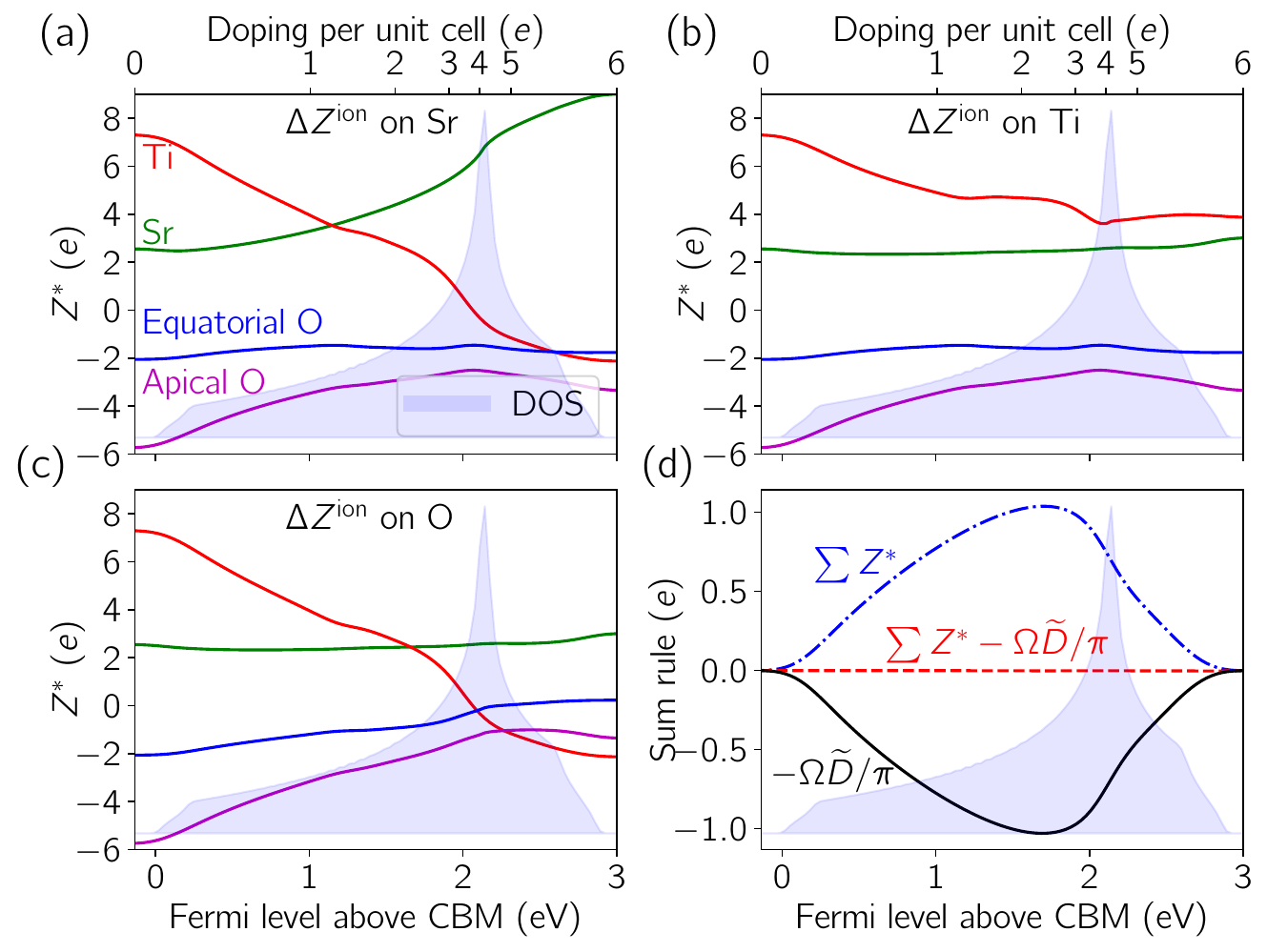}
\caption{\label{STOtot} Nonadiabatic Born effective charges of the sublattices in SrTiO$_3$ versus doping above the conduction-band minimum on the (a) Sr, (b) Ti, or (c) O sublattice. (d) Demonstrates the naBEC sum rule, where the sublattice sum of the BECs cancels with the Drude weight. Blue shaded region is the density of states of the Wannierized Ti $t_{2g}$ manifold.
}
\end{figure}

In Fig.~\ref{STOtot}(a)-(c) we plot the naBECs for the sublatticies in STO with doping. 
The oxygens are labelled as ``equatorial'' (on faces parallel to $\hat{i}$ for BEC element $Z^*_{ii}$) and ``apical'' (on faces perpendicular to $\hat{i}$).
Different trends in $Z^*$ result from different choices of sublattice for chemical doping,
suggesting significant qualitative changes in frequencies of
polar-phonon modes depending on the sublattice that is doped versus
the sublattice(s) involved in the phonon
displacements. Fig.~\ref{STOtot}(a) corresponds to doping on the Sr
site, e.g., by La substitution; in this case, the Sr $Z^*$ increases
significantly to compensate the decrease of the Ti $Z^*$, resulting from the fact that the electrons are
doped into the Ti $3d$ orbitals;this effect is significantly reduced if we consider doping on the Ti site in panel (b) (e.g., by Nb substitution). For doping on the O site (e.g., by the formation of O vacancies), the decrease in the Ti  $Z^*$ is now partially compensated by the increase in $Z^*$ for the O sublattices.

In Fig.~\ref{STOtot}(d), we see very similar behavior to Fig.~\ref{SnS2_phi}(b), i.e., the sublattice sum of the BECs increasingly deviates from zero with doping towards the middle of the conduction-band manifold, and then decreases. As with \sns{}, the naBEC sum rule is accurately satisfied.

In conclusion, we have demonstrated that nonadiabatic Born effective charges are well defined in metals with time-reversal symmetry, and generalized the acoustic sum rule to the full nonadiabatic regime, where  the sublattice sum of the nonadiabatic Born effective charges equals the Drude weight. 
The rigorous understanding and first-principles description of nonadiabatic Born effective charges provided by this work opens up several future directions for study. For example, it will allow quantitative predictions of plasmon-phonon coupling, and the implications for transport in doped semiconductor devices such as transparent conductors.
Direct comparison with experimental probes such as infrared or Raman scattering will serve to validate the theory and aid in materials characterization. 
It may also shed light on the physics of ferroelectric metals, possibly providing additional tools to quantify the amplitude of the polar lattice distortion.

\begin{acknowledgments}
We thank D. Vanderbilt and R. Resta for discussions and insightful comments on the manuscript. CED
acknowledges support from the National Science Foundation under Grant No. DMR-1918455.
SC
acknowledges support from the National Science Foundation under Grant No. DMR-1848074.
MS acknowledges support from the European
 Research Council (ERC) through Grant ``MULTIFLEXO'' No. 724529; from Ministerio de Economia, Industria y Competitividad (MINECO-Spain) through Grants  No. PID2019-108573GB-C22  
 and No. CEX2019-000917-S; and from Generalitat de Catalunya (Grant No. 2017 SGR1506). The Flatiron Institute is a division of the Simons Foundation.
\end{acknowledgments}

\bibliography{flexo}

\end{document}


\title{Supplemental Material: Nonadiabatic Born effective charges in metals}
\author{Cyrus E. Dreyer}
\affiliation{Department of Physics and Astronomy, Stony Brook University, Stony Brook, New York, 11794-3800,
  USA}
\affiliation{Center for Computational Quantum Physics, Flatiron Institute, 162 5th Avenue, New York, New York 10010,
USA}
\author{Sinisa Coh}
\affiliation{Materials Science and Mechanical Engineering, University of California Riverside, CA 92521, USA}
\author{Massimiliano Stengel}
\affiliation{Institut de Ci\`{e}ncia de Materials de Barcelona (ICMAB-CSIC), Campus UAB, 08193 Bellaterra, Spain}
\affiliation{ICREA-Instituci\'{o} Catalana de Recerca i Estudis Avan\c{c}ats, 08010 Barcelona, Spain}
\date{\today}

\begin{abstract}

\end{abstract}

\maketitle

\section{Details of DFPT formalism \label{sm:form}}

We will perform our derivation and ultimately our calculations in the
framework of planewave/pseudopotential Kohn-Sham \cite{Kohn1965} DFT. Thus
$\psi_\nk(\textbf{r})=u_\nk(\textbf{r})
e^{i\textbf{k}\cdot\textbf{r}}$ corresponds to the single-particle
wavefunction of band $n$ and wavevector \textbf{k} in the
first Brillouin zone that is the solution of the Kohn-Sham equation
with the single particle Hamiltonian
\begin{equation}
\hat{\mathcal{H}}=  \hat{T}+\hat{V}_\hxc[\rho]+\hat{V}_{\text{ext}},
\end{equation}
where $\hat{T}$ is the single-particle kinetic energy,
$\hat{V}_{\text{Hxc}}[\rho]$ denotes the Hartree and
exchange-correlation (Hxc) potential, which is a functional of the density
$\rho$, and $\hat{V}_{\text{ext}}$ is the external potential that
includes the pseudopotential operators (local and nonlocal). We will
often use the ``cell-periodic'' version of operators:
$\hat{\mathcal{O}}_{\textbf{k}}=e^{-i\textbf{k}\cdot\textbf{r}}\hat{\mathcal{O}}e^{i\textbf{k}\cdot\textbf{r}}$
which operate on the cell-periodic functions $u_\nk(\textbf{r})$.

\subsection{Generalized susceptibility}

We begin with the generalized  
susceptibility introduced in the main text, and partition it in the following way 
\begin{equation}
\label{chi}
\chi^{\text{na}}_{\lambda_1\lambda_2}(\omega,\textbf{q})=\chi^{\text{geom}}_{\lambda_1\lambda_2}(\textbf{q})+\chi^{\text{Kubo}}_{\lambda_1\lambda_2}(\omega,\textbf{q}).
\end{equation}
The ``geometric'' contribution only depends on the unperturbed density operator, $\hat{\rho}$, and is independent of $\omega$. Taking immediately the limit of $\textbf{q}\rightarrow0$, 
\begin{equation}
  \label{rho0}
\chi^{\text{geom}}_{\lambda_1\lambda_2}= 
 \text{Tr} \, \left( \hat{H}^{\lambda_1\lambda_2}_{\bf k} \,  \hat{\rho} \right), \quad 
 \hat{\rho}_{\bf k} = \sum_n  \vert u_\nk\rangle  f_\nk\langle u_\nk\vert,
\end{equation}
where $\hat{H}_{\textbf{k}}^{\lambda_1\lambda_2}=\partial^2
\hat{H}_{\textbf{k}}/\partial\lambda_1\partial\lambda_2$ is the second derivative of the external potential (i.e., not containing the SCF part), and $f_\nk$ is the occupation of band $n$ at $k$-point
\textbf{k}. Note that the trace implicitly consists of a sum over bands and Brillouin-zone average,
\begin{equation}
    {\rm Tr} \left( \hat{A} \hat{B} \right) = 
    \int [d^3 k] \sum_{mn} 
 \langle u_{m{\bf k}} | \hat{A}_{\bf k} | u_{n{\bf k}} \rangle 
 \langle u_{n{\bf k}} | \hat{B}_{\bf k}| u_{m{\bf k}} \rangle.
\end{equation}
The second term in Eq.~(\ref{chi}) can be written as 
\begin{equation}
  \label{kubo}
  \chi^{\text{Kubo}}_{\lambda_1\lambda_2}(\omega,\textbf{q}) = 
   \text{Tr} \, \left[  
   \hat{H}^{\lambda_1 \dagger}_{\textbf{k},\textbf{q}} \,
   \hat{\rho}^{\lambda_2}(\omega, {\bf q} ) \right],
\end{equation}
where the first-order density matrix is written as
a double sum over states
\begin{equation}
  \label{rholam}
  \begin{split}
      \hat{\rho}^{\lambda}_{\bf k} (\omega,\textbf{q}) & = \lim_{\eta \rightarrow 0} \sum_{nm} \bar{f}_{nm\bfk}(\omega + i\eta,{\bf q})
    \vert\umkq\rangle\langle\umkq\vert
     \hat{\mathcal{H}}^{\lambda}_{\textbf{k},\textbf{q}}(\omega)\vert\unk\rangle
    \langle \unk \vert,
\end{split}
\end{equation}
with
\begin{equation}
\label{eq:overlinef}
\bar{f}_{nm\bfk}(z,{\bf q}) =
  \frac{f_\nk-f_\mkq}{\epsilon_\nk-\epsilon_\mkq+z}.
\end{equation}
Note that the first-order Hamiltonian becomes $\omega$-dependent via the charge
self-consistency once the local fields are incorporated,

\begin{equation}
\hat{\mathcal{H}}^\lambda(\omega)=\hat{H}^\lambda+\hat{V}_\hxc^\lambda(\omega)
\end{equation}
where $\hat{H}^\lambda$ is the kinetic energy and external potential perturbation, and
$\hat{V}_{\text{Hxc}}^\lambda$ is the potential response containing
self-consistent fields (SCF) that depends on the first-order density,
$\rho^{\lambda}_{\bf q}(\textbf{r},\omega) =    
   \langle {\bf r} | \,
   \hat{\rho}^{\lambda_2}(\omega, {\bf q} ) |{\bf r}\rangle $
as
\begin{equation}
\hat{V}_{\text{Hxc}}^\lambda(\textbf{r},\omega)=\int d^3r^\prime K_\hxc(\textbf{r},\textbf{r}^\prime)\rho^{\lambda,\omega}(\textbf{r},\omega),
\label{vhxc}
\end{equation}
where $K_\hxc$ is the Hxc kernel, defined as the variation of the SCF potential with respect to a charge-density perturbation.

 \subsection{Adiabatic regime \label{sm:ad}}

If we take $\omega \rightarrow 0$, the intraband contribution in 
Eq.~(\ref{kubo}) remains finite, and
\begin{equation}
\label{eq:overlinefnn}
    \overline{f}_{nn\bfk} \rightarrow
    \frac{df(\epsilon)}{d\epsilon}\Bigg\vert_{\epsilon=\epsilon_\nk}.
\end{equation}
Then, for a $\textbf{q}=0$ perturbation, we obtain the \emph{adiabatic} response,
\begin{equation}
  \label{KuboAd}
\chi^{\text{ad}}_{\lambda_1\lambda_2} = %
\text{Tr}\left[ \frac{\partial}{\partial \lambda_2} 
 \left( \hat{\rho} \hat{H}^{\lambda_1} \right)\right].
\end{equation}
Note that 
$\chi^{\text{ad}}_{\lambda_1\lambda_2}$ is written as
a total derivative with respect to one of the perturbations.
This means that, if either the $\lambda_1$ or $\lambda_2$ is a vector potential field, the corresponding perturbation reduces to a partial derivative with respect to $k_\alpha$, and its BZ average vanishes.
Physically, the fact that $\chi^{\text{ad}}_{A_\alpha A_\beta}$ vanishes is a direct consequence of the $f$-sum rule; the result $\chi^{\text{ad}}_{A_\alpha \tau_{\kappa\beta}}=0$, on the other hand, tells us 
that a static atomic displacement cannot produce a steady current.
$\chi^{\text{ad}}_{\tau_{\kappa\alpha} \tau_{\kappa'\beta}}$ does not vanish: it is the electronic contribution to the adiabatic force-constant matrix, as it is calculated in most linear-response DFT packages
\cite{Baroni2001,Gonze1997,GonzeLee1997}.

 \subsection{Nonadiabatic response}

 When considering, e.g., 
optical phonons in polar crystal lattices, or the 
response to electromagnetic radiation,
it is appropriate to reverse the order of the $\textbf{q}\rightarrow0$ and $\omega\rightarrow 0$ limits.
The intraband contributions to Eq.~(\ref{kubo}) is suppressed in the $\textbf{q}\rightarrow0$ limit, 
thus, we obtain a \emph{nonadiabatic} response function 
as the \emph{interband} part of the adiabatic one, i.e., by replacing $\hat{\rho}^{\lambda_2}$ in Eq.~(\ref{kubo}) with 
\begin{equation}
  \begin{split}
  \label{sm:pinter}
  \hat{\rho}^{\lambda_2,\text{inter}}_{\textbf{k}}
  &=
  \sum_{n\neq m} \overline{f}_{nm}^{\bf k}
  \vert\umk\rangle\langle\umk\vert
  \hat{\mathcal{H}}^{\lambda_2}_{\textbf{k}}
  \vert\unk\rangle\langle\unk\vert.
\end{split}
\end{equation}
Note that, in principle, the intraband contribution to the first-order charge density should be removed as well, i.e., the adiabatic scattering potential $\hat{\mathcal{H}}^{\lambda_2}$ should be replaced with $\hat{\mathcal{H}}^{\lambda_2,\text{inter}}=\hat{H}^{\lambda_2}+\hat{V}_\hxc^{\lambda_2,\text{inter}}$,
and $\hat{V}_\hxc^{\lambda_2,\text{inter}}$ is defined from
$\rho^{\lambda_2,\text{inter}}({\bf r})$ via Eq.~(\ref{vhxc}).
While taking this extra step would be desirable, as it would lead to a fully self-consistent computational setup, it would also substantially complicate the implementation; therefore, in our calculations we have retained the adiabatic phonon perturbation, $\hat{\mathcal{H}}^{\lambda_2}$, for simplicity.
We believe, in fact, that our calculations are unaffected by such a simplifying assumption, as we shall clarify in the following.

At small (but finite) ${\bf q}$, correctly treating the intraband contribution to the density in the SCF cycles may be important. Whether free carriers participate (full adiabatic density) or do not participate (nonadiabatic interband-only density) to screening a phonon perturbation has obviously a crucial impact on the macroscopic electric fields that are produced, e.g., by a long-wavelength LO mode. 
Right at the zone center, however, and under the assumption of short-circuit electrical boundary conditions (as required by the definition of the Born dynamical charge tensor), neglecting the nonadiabatic correction to the SCF potential is much safer, since it may only lead to a small discrepancy in the first-order potential that averages to zero over the cell.
Furthermore, in presence of space-inversion symmetry, infrared-active modes (i.e., the only ones carrying a nonzero dynamical charge) do not couple with Fermi-level shifts, and hence their intraband contribution to the density response should vanish.
This indicates that our calculations (all performed on centrosymmetric crystals) should not be affected by this issue.

 \subsubsection{Drude weight}

Consider the case where both $\lambda_1$ and $\lambda_2$ are
components of the vector potential. The total density response to a translation in \textbf{k}-space vanishes, so the interband contribution to the density response is minus the intraband one. Also, the density response, and thus the SCF contributions vanish if TRS is present (since  
$\overline{f}_{n\nk}$ and $\vert\unk(\textbf{r})\vert^2$ are even under
$\textbf{k} \rightarrow -\textbf{k}$, while
$\langle\unk\vert\hat{H}^{A_\alpha}_{\textbf{k}}\vert\unk\rangle$ is
odd). Thus, the response reduces to the
familiar Drude expression (e.g, see Ref.~\onlinecite{Resta2018}),
\begin{equation}
  \begin{split}
    \label{drude}
    \chi^{\text{na}}_{A_\alpha A_\beta}&=-\int[d^3k]
    \overline{f}_{n\nk}\langle\unk\vert\hat{H}^{A_\alpha}_{\textbf{k}}\vert\unk\rangle\langle\unk\vert\hat{H}^{A_\beta}_{\textbf{k}}\vert\unk\rangle
    \\
    &=-\int[d^3k]
    \overline{f}_{n\nk}\frac{\partial\epsilon_\nk}{\partial k_\alpha}\frac{\partial\epsilon_\nk}{\partial k_\beta}=\frac{\Omega}{\pi}D_{\alpha\beta},
    \end{split}
\end{equation}
where $\Omega$ is the cell volume. $D_{\alpha\beta}$ is the ``Drude
weight.''
This corresponds to the optical conductivity
multiplied by $i\omega$. The Drude weight is nonzero in all metals,
regardless of crystal symmetry, which is seen by realizing that the
square of the Fermi velocities in Eq.~(\ref{drude}) is even under
$\textbf{k}\rightarrow-\textbf{k}$ even if the crystal has TR and
inversion symmetry.

As discussed in the main text, we will also consider a ``modified'' version of the Drude weight, given by
\begin{equation}
  \begin{split}
    \label{drudemod}
    \frac{\Omega}{\pi}\widetilde{D}_{\alpha\beta}&=-\int[d^3k]
    \overline{f}_{n\nk}\langle\unk\vert\hat{H}^{A_\alpha}_{\textbf{k}}\vert\unk\rangle\langle\unk\vert\hat{p}_{\beta\textbf{k}}\vert\unk\rangle
        \end{split}
\end{equation}
where $\hat{p}_{\beta,\textbf{k}}=-i\nabla_\beta+\hat{k}_\beta$ is the canonical momentum operator. 

\subsubsection{Born effective charges \label{sm:naBEC}}

We will now derive the nonadiabatic response at first order in the velocity, valid for the optical conductivity and the Born effective charges.
The derivation rests on the following identity
\begin{equation}
    \frac{\Delta f}{\Delta \epsilon + z} =
    \frac{\Delta f}{\Delta \epsilon} \left( 1 - \frac{z}{\Delta \epsilon + z}
 \right).
\end{equation}
The first term in the round brackets does not depend on frequency; summed with the geometric term it yields the adiabatic response, which vanishes both for the current--current and the current--force response functions.
From the second term in the round brackets we readily obtain, via Eq.~(1) of the main text, the established formula for the optical conductivity (i.e., Eq.~(25) of Ref.~\onlinecite{Allen}).
Similarly, Eq.~(2) of the main text yields
the electronic contribution to the Born charges as \cite{Binci2020,Bistoni2019}
  \begin{equation}
    \begin{split}
      \label{NAelBEC}
  Z^{(\alpha)}_{\kappa \beta}(\omega + i\eta),
    &= 
    - \frac{1}{i} \int [d^3 k]
    \sum_{n m}
    \frac{\overline{f}_{n\mk}}{\epsilon_\nk-\epsilon_\mk + z}
    \langle\unk\hat{H}_{\textbf{k}}^{k_\alpha\dagger}\vert\umk\rangle\langle\umk\vert\hat{\mathcal{H}}^{\tau_\beta}_{\textbf{k}}(z)\vert\unk\rangle,
    \end{split}
  \end{equation}
  where $\hat{H}_{\textbf{k}}^{k_\alpha}$ is the velocity operator and $\tau_{\kappa\beta}$ indicates the displacement of the sublattice $\kappa$ along the Cartesian direction $\beta$. After taking the $\omega \rightarrow 0$ limit, we readily obtain the expression for the naBEC given by Eq.~(7) of the main text. In an insulator, Eq.~(7) reduces to the usual expressions for the adiabatic BEC.

 \subsection{Sums over empty states
 \label{sm:comp}}
 
If we were to calculate the naBEC using Eq.~(7), then we would have to perform a convergence over empty states. In order to avoid this, we will use the same approach as is common in DFPT, casting the problem in terms of obtaining first-order wavefunctions via solving a Sternheimer equation. 
To this end, we shall assume that the active subspace of states that are treated explicitly in the calculation spans the lowest $M$ orbitals with occupation numbers different from zero.
Then, we can decompose Eq.~(7) into a double sum over the first $M$ states plus two sums over $1,\ldots,M$ and $M+1,\ldots$,
\begin{equation}
  \begin{split}
    \lim_{\eta \rightarrow 0^+} &
  \int [d^3 k]
    \sum_{n \leq M, m \leq M}
    \frac{\overline{f}_{n\mk}}{\epsilon_\nk-\epsilon_\mk + i \eta}
    \langle\unk | \hat{H}_{\textbf{k}}^{k_\alpha}
    \vert\umk\rangle \langle\umk\vert
    \hat{\mathcal{H}}^{\tau_{\kappa \beta}, \rm inter}_{\textbf{k}}\vert\unk\rangle  \\
   & +  \int [d^3 k]
    \sum_{n \leq M, m \geq M+1}
    \frac{f_\nk}{(\epsilon_\nk-\epsilon_\mk)^2}
    \langle\unk | \hat{H}_{\textbf{k}}^{k_\alpha}
    \vert\umk\rangle \langle\umk\vert
    \hat{\mathcal{H}}^{\tau_{\kappa \beta}, \rm inter}_{\textbf{k}}\vert\unk\rangle \\
   & -   \int [d^3 k]
    \sum_{m \leq M, n \geq M+1}
    \frac{f_\mk}{(\epsilon_\nk-\epsilon_\mk)^2}
    \langle\unk | \hat{H}_{\textbf{k}}^{k_\alpha}
    \vert\umk\rangle \langle\umk\vert
    \hat{\mathcal{H}}^{\tau_{\kappa \beta}, \rm inter}_{\textbf{k}}\vert\unk\rangle  
    \end{split}
\end{equation}
The second and third terms can be conveniently rewritten as
a Berry curvature in parameter space, so the naBEC can be written as
\begin{equation}
\label{Zimp}
\begin{split}
 Z^*_{\alpha,\kappa\beta}&=Z^{\text{ion}}_{\kappa}\delta_{\alpha\beta} - \text{Im} \Bigg[\int [d^3 k]\sum_{n \leq M} f_\nk \left(
    \langle\unk^{k_\alpha} \vert 
    \unk^{\tau_{\kappa \beta}, \rm inter} \rangle 
 -    \langle\unk^{\tau_{\kappa \beta}, \rm inter}
   \vert \unk^{k_\alpha} \rangle
  \right)\\
      &+\lim_{\eta \rightarrow 0^+}
  \int [d^3 k]
    \sum_{n \leq M, m \leq M}
    \frac{\overline{f}_{n\mk}}{\epsilon_\nk-\epsilon_\mk + i \eta}
    \langle\unk | \hat{H}_{\textbf{k}}^{k_\alpha}
    \vert\umk\rangle \langle\umk\vert
    \hat{\mathcal{H}}^{\tau_{\kappa \beta}, \rm inter}_{\textbf{k}}\vert\unk\rangle \Bigg]
    \end{split}
\end{equation}
where the first-order wavefunctions are 
defined via 
\begin{equation}
  \label{stern}
(\hat{\mathcal{H}}_\bfk+a\hat{P}^M_\bfk-\epsilon_\nk)\vert u_\nk^{\lambda}\rangle=-\hat{Q}^M_\bfk\hat{\mathcal{H}}^\lambda_\bfk\vert u_\nk\rangle,
\end{equation}
where $\hat{P}^M_\bfk-\sum_n^M\vert u_\nk\rangle\langle u_\mk\vert$ and
$\hat{Q}^M_\bfk=1-\hat{P}_\bfk$ are projectors inside and outside the active
space, and $a$ is a constant that needs to be larger than the total
bandwidth of the active space. Eq.~(\ref{stern}) has the same form as the Sternheimer equation solved
presently in DFPT implementations
\cite{Gonze1997,GonzeLee1997,Baroni2001}, except that
$\hat{P}^M_\bfk$ does not correspond to the ground-state density operator, but instead the projector over an ``active space'' $M$, which is chosen to be large enough to
encompass all states such that the occupation of the $M$th band
vanishes. We will show in Sec.~\ref{sm:comp} that the choice of $M$ does not influence the final result.

\section{Computational approach and convergence \label{sm:comp2}}

\subsection{Computational parameters}
We implemented the methodology for calculating the naBEC and DW in the {\sc abinit} package~\cite{Abinit_1}, taking advantage of the DFPT implementation for calculating the (static) response to atomic displacements and infinitesimal electric fields \cite{Gonze1997,GonzeLee1997}. In all cases, $\eta$ is fixed to be $10^{-5}$ Ha. Optimized Vanderbilt norm-conserving pseudopotentials
\cite{Hamann2013} taken from {\sc pseudo-dojo} are used for all
calculations.

\begin{figure}
\includegraphics[width=\columnwidth]{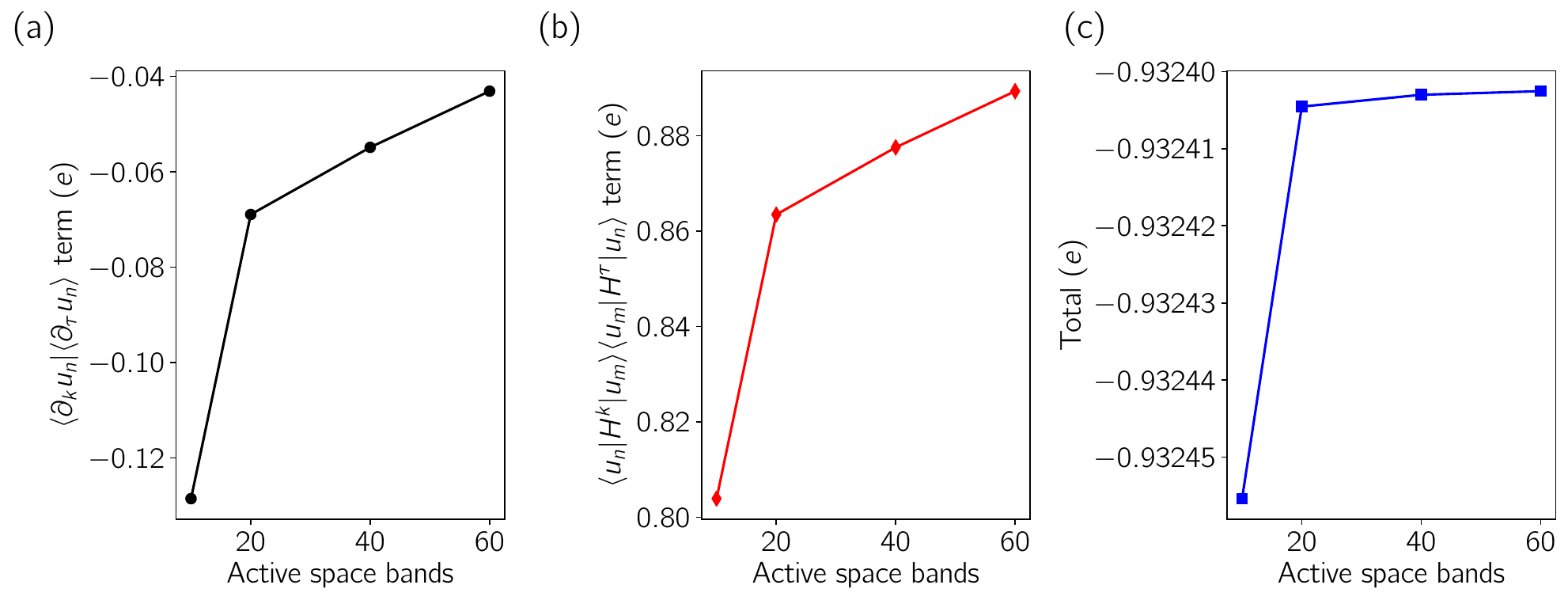}
\caption{\label{Al_mbands} For cubic cell of FCC Al,  the dependence on choice of number of bands in the ``active space'' $M$ of (a) the second term in Eq.~(\ref{Zimp}), (b) the third term in Eq.~(\ref{Zimp}), (c) the total electronic contribution to the nonadiabatic Born effective charge. }
\end{figure}

For Al, we use the PBE \cite{pbe} generalized-gradient approximation
exchange-correlation functional, as it provides better structural properties compared to LDA (cubic lattice parameter of 7.63 Bohr). We use Fermi-Dirac smearing with a ``temperature'' of 0.005 Ha. As mentioned in the main text, $k$-meshes up to $28\times28\times28$ were used. For the calculation of the naBEC, we used an active space of $M=40$ bands; we show in Fig.~\ref{Al_mbands} that the total naBEC does not depend on such choice, as long as enough bands are included such that the occupation of the highest band is negligible ($M\geq20$ for the conventional cubic cell).

\begin{figure}
 \includegraphics[width=\columnwidth]{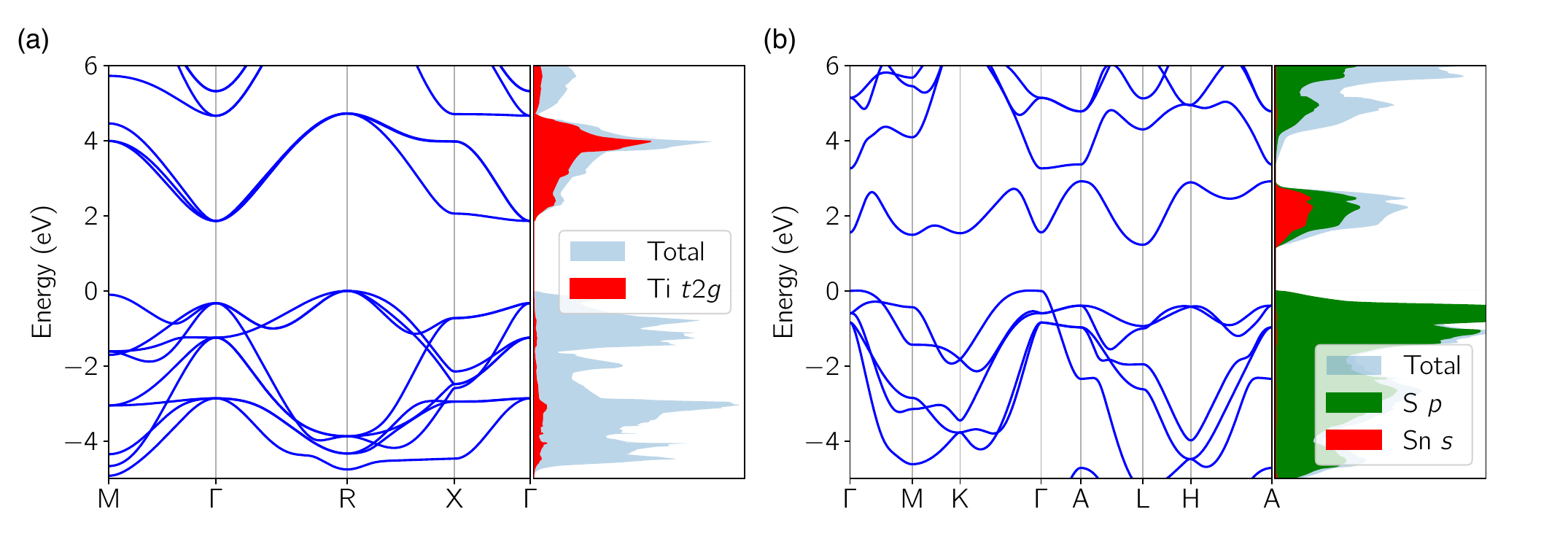}
  \caption{Bandstructure and projected density of states (DOS) for (a) $Pm\overline{3}m$ SrTiO$_3$ and (b) SnS$_2$. Zero energy for each is set to the top of the valence band.}\label{bulk_bs_dos}
\end{figure}

For cubic ($Pm\overline{3}m$) SrTiO$_3$ (STO) and SnS$_2$, the local density approximation (LDA) functional parametrized in Ref.~\onlinecite{Perdew1992} was used (as they provide better structural properties than GGA for these materials). The relaxed lattice parameter(s) for STO was $7.29$ Bohr and for \sns{} were $a=6.87$ and $c=10.78$ Bohr. Gaussian smearing was used with a width of 0.005 Ha, and a ``coarse'' $k$-mesh of $8\times8\times8$ was used for the undoped calculations for STO and 0.001 Ha and $12\times12\times6$ for \sns{}. (The slightly larger smearing for STO was in order to smooth fluctuations due to the sharp features in the Fermi surface, see Sec. \ref{sm:conv}) The linear response quantities necessary for the calculation of the naBEC and DW with electron doping were Wannier interpolated onto a ``fine'' mesh in $k$ space as discussed in Sec.~\ref{FOwannier}. 

\subsection{First-order Wannier functions for interpolation \label{FOwannier}}

\begin{figure}
\includegraphics[width=\columnwidth]{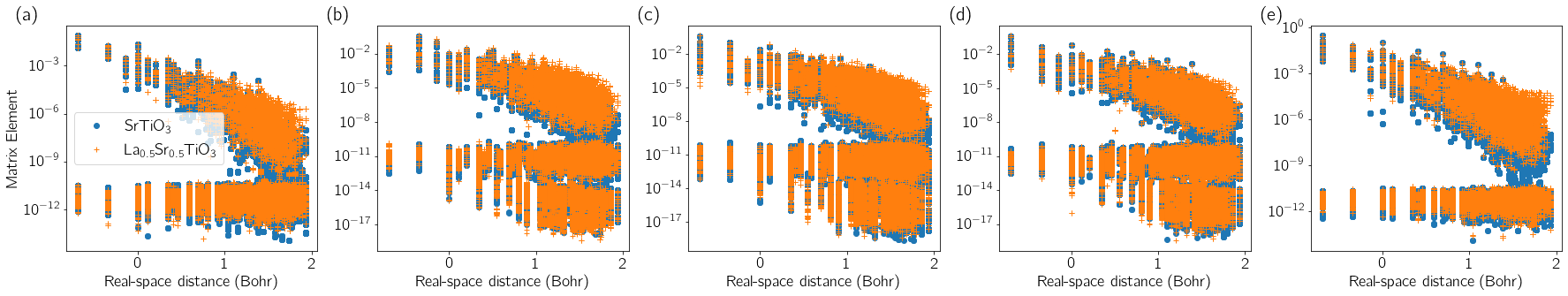}
\caption{\label{Decays} Decay of the real-space matrix elements of (a) the ground state Hamiltonian, (b) atomic displacement perturbations, (c) velocity operator, (d) momentum operator, and (e), the matrix elements involved in the second term in Eq.~(\ref{Zimp}). Blue circles are for undoped STO, orange crosses are doped with La (0.5$e$ per cell) via the virtual crystal approximation.}
\end{figure}

In order to achieve the meshes in $k$ space necessary to converge the naBEC and DW for the doped semiconductors STO and \sns{} (see the next section for discussion of the doping), we performed a Wannier interpolation of the quantities in Eqs.~(\ref{Zimp}), and (\ref{drude}). Maximally-localized
Wannier functions are generated using the {\sc Wannier90} \cite{wannier90} interface with {\sc abinit} for the valence bands as well as the three lowest-lying conduction bands for STO [Fig.~\ref{bulk_bs_dos}(a)], and single lowest-lying conduction band for \sns{} [Fig.~\ref{bulk_bs_dos}(b)]. In order to ensure that the a consistent gauge is used, DFPT calculations are initiated from the same  ground-state wavefunctions as used for the Wannierization. The unitary transformations that produced the Wannier functions of the ground-state Hamiltonian are used to interpolate the first-order Hamiltonian and wavefunction derivatives (a similar strategy as employed in Refs.~\onlinecite{Ge2015} and \onlinecite{Lihm2021}) onto a fine $k$-mesh. In order to confirm that this procedure is providing a valid localized basis for interpolation, we plot the decay of the ground state and first-order Hamiltonian matrix elements for STO in real space in Fig.~\ref{Decays} (a similar behavior is observed for \sns{}). For the results in the main text, we use a fine mesh of $100\times100\times100$ for STO, and $128\times128\times64$ for \sns{}.

\subsection{Electron doping}

\begin{figure}
\includegraphics[width=\columnwidth]{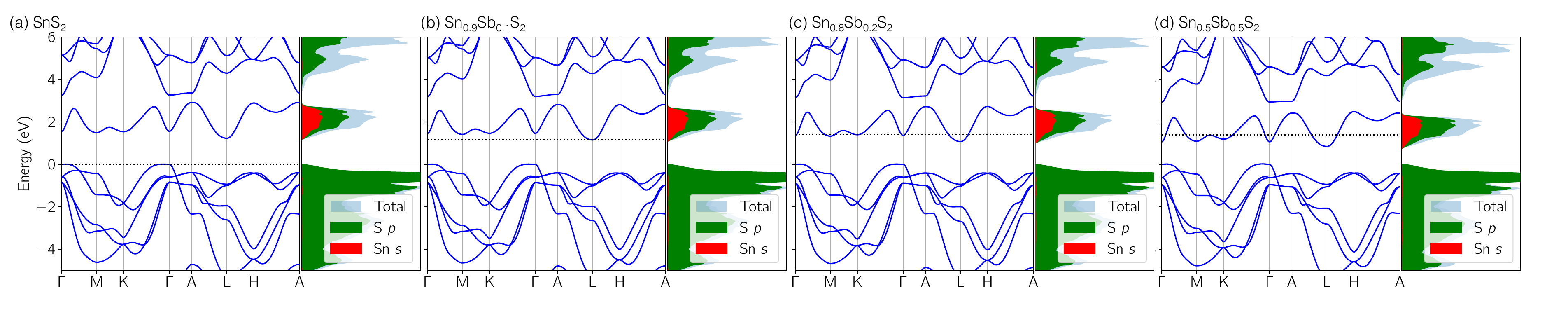}
\caption{\label{Sns2_doping} Band structure using the virtual crystal approximation for Sb$_x$Sn$_{1-x}$S$_2$ with various Sb concentrations. Dotted line is the Fermi level, and energy zero is set to the top of the valence band in all cases.}
\end{figure}

\begin{figure}
\includegraphics[width=\columnwidth]{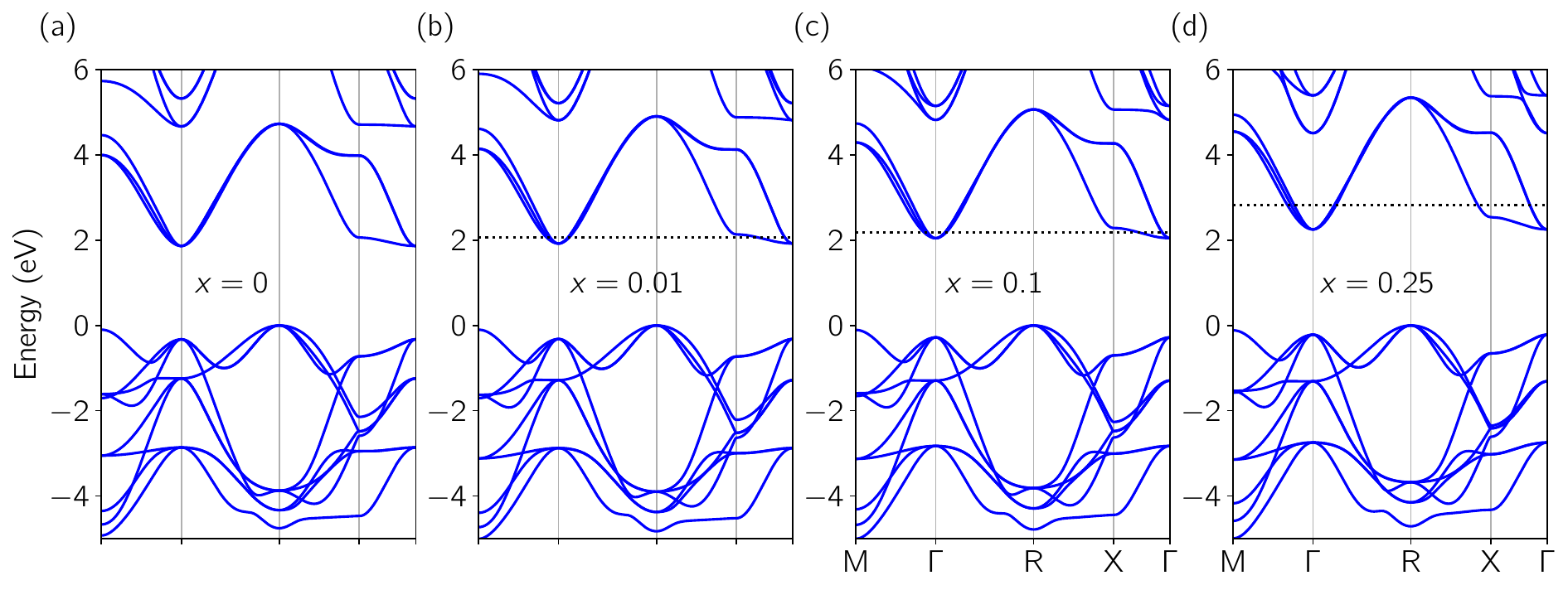}
\caption{\label{LSTO_bs} Band structure using the virtual crystal approximation for \lsto with various La concentrations. Dotted line is the Fermi level, and energy zero is set to the top of the valence band in all cases.}
\end{figure}

For Figs.~2 and 3 in the main text, \sns{} and STO were doped via the rigid-band approximation, i.e., only the occupation factors were changed with Fermi level. For STO, doping was performed across the entire Ti $3d$ $t_{2g}$ conduction band manifold [see Fig.~\ref{bulk_bs_dos}(a)], while for \sns{}, the doping was performed across the single conduction band [see Fig.~\ref{bulk_bs_dos}(b)], which has S $3p$ and Sn $5s$ character.

We can explore the accuracy of this approximation by comparing with explicit doping via the virtual-crystal approximation (VCA). To do this, the Sn (Sr) pseudopotential was alchemically mixed with a Sb (La) pseudopotential, with all structural properties kept fixed. The same Wannier interpolation procedure  (Sec.~\ref{FOwannier}) was used in these cases as in the rigid-band calculations. In Figs.~\ref{Sns2_doping} and \ref{LSTO_bs}, we show that the band structures do not change significantly with doping via VCA (as long as the structural parameters are fixed). From, e.g., Fig.~\ref{Sns2_doping}(d), we do see some subtle changes at large doping for \sns{}, i.e., it can be seen from the DOS that as the S $p$/Sn $s$ conduction band is filled, the splitting to the S $p$ valence band decreases slightly. 

In Fig.~\ref{Decays} we compare the ground state (panel a) and first-order matrix elements that make up the naBEC and Drude weight (panels b-e) calculated for undoped STO (blue circles) and VCA La$_{0.5}$Sr$_{0.5}$Ti$_3$ (orange crosses). Overall, there is good agreement between the two cases, even for such a large doping; the spatial decay is somewhat reduced for the VCA doped case, which is likely due to slightly less localized Wannier functions in the metallic case. 

\begin{figure}
\includegraphics[width=0.9\columnwidth]{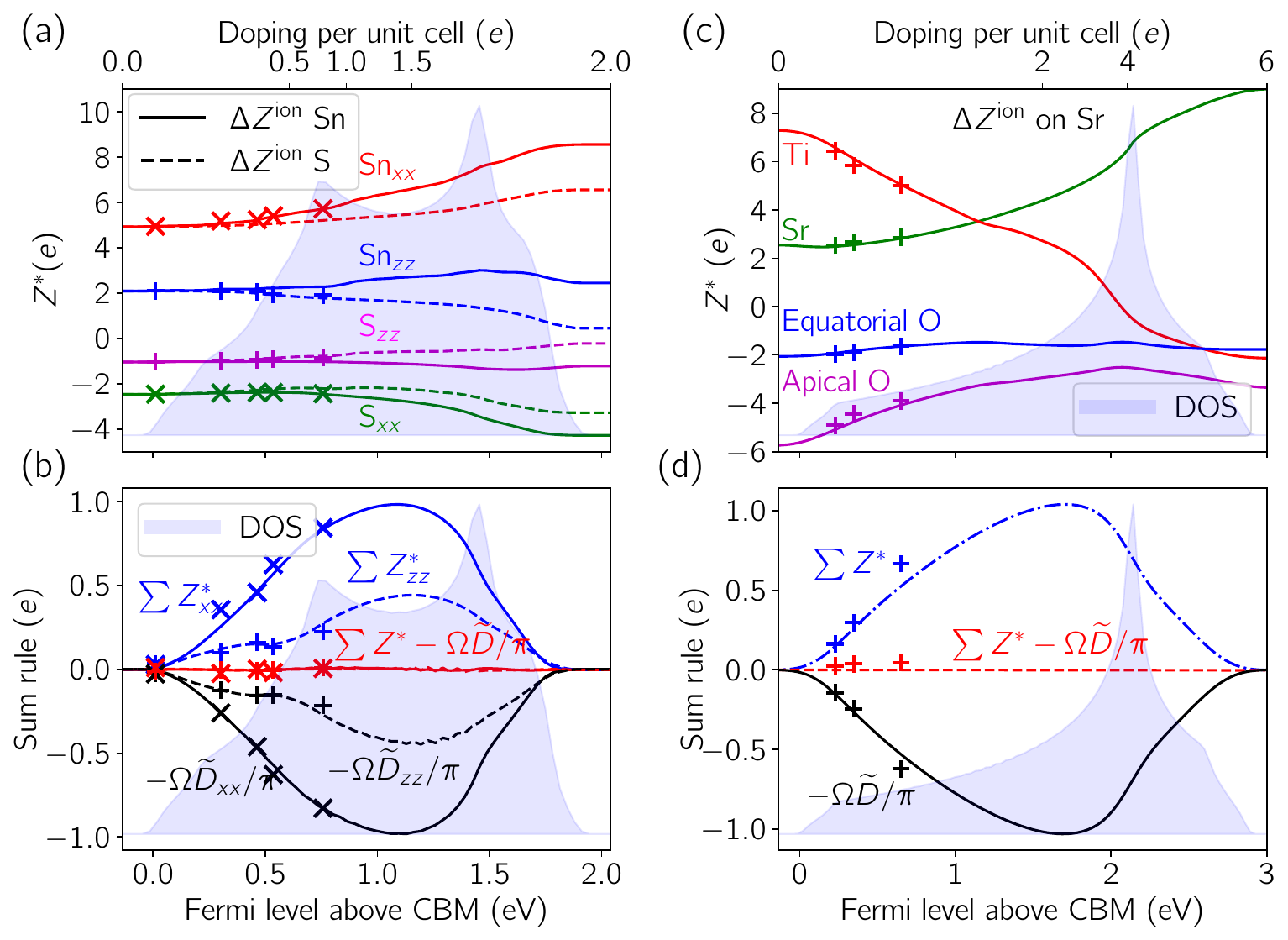}
\caption{\label{fig:VCA} Comparison of explicit doping via the virtual-crystal approximation (+'s and x's) and the rigid-band approximation (curves, same as Figs.~2 and 3 in the main text). Points in (a) and (b) are obtained by alchemically mixing the Sn pseudopotential with Sb, and in (c) and (d) mixing Sr with La. }
\end{figure}

In Fig.~\ref{fig:VCA} we compare the naBECs and sum rule for \sns{} and STO calculated with the VCA (points) and rigid-band approximations (curves, same as Figs.~2 and 3 in the main text). We can see that the agreement is excellent at low doping. At higher doping there are some quantitative differences. In the case of \sns{}, there are some small deviations at 50\% Sb doping in the $z$ components of the naBEC; for STO at 50\% La doping, the total naBEC and Drude weight are slightly larger than for the rigid-band approximation. However, crucially, the naBEC sum rule is accurately satisfied in both cases, and thus does not depend on how we apply the doping. The increased scatter in the VCA points does indicate more challenging convergence when the dopant electrons are explicitly included.

\subsection{Convergence of nonadiabatic Born effective charges and Drude Weight \label{sm:conv}}

\begin{figure}
\includegraphics[width=\columnwidth]{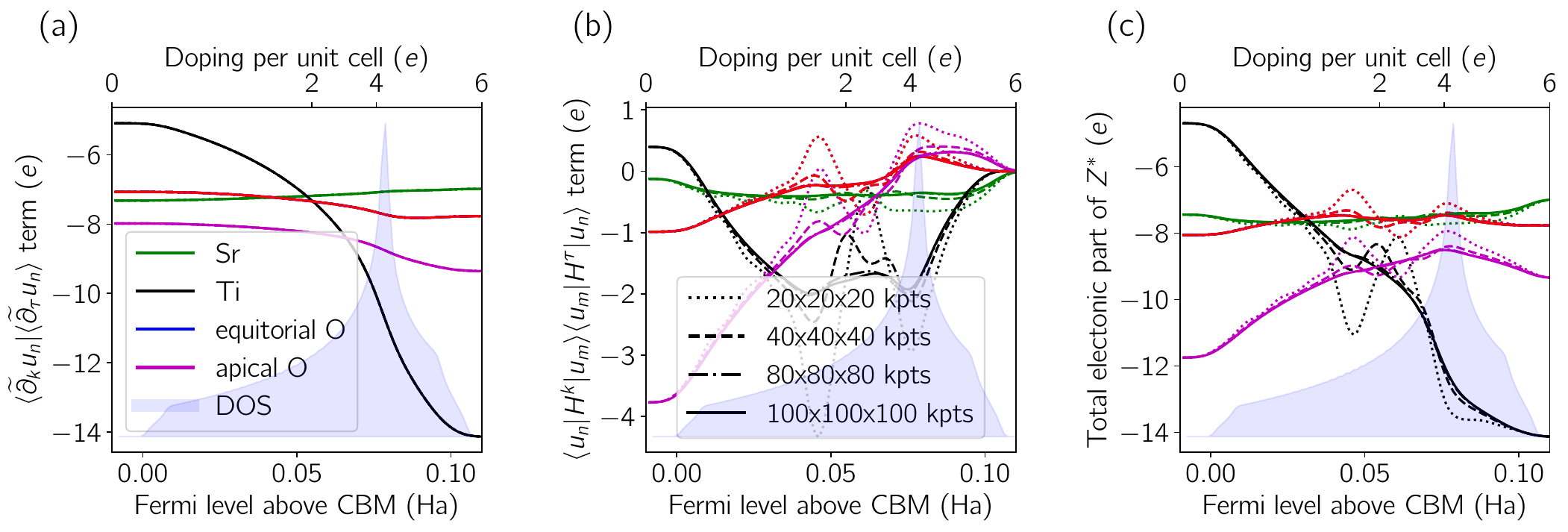}
\caption{\label{BEC_STO_kpts} Convergence of the electronic part of the naBECs of STO with interpolated $k$ mesh. (a) is the second term in Eq.~(\ref{Zimp}), (b) is the third term in Eq.~(\ref{Zimp}), and (c) is the total electronic contribution to the nonadiabatic Born effective charge. The density of states of the Ti $t2g$ manifold is superimposed, and the thermal smearing is set to $\sigma=0.005$ Ha.  }
\end{figure}

\begin{figure}
\includegraphics[width=\columnwidth]{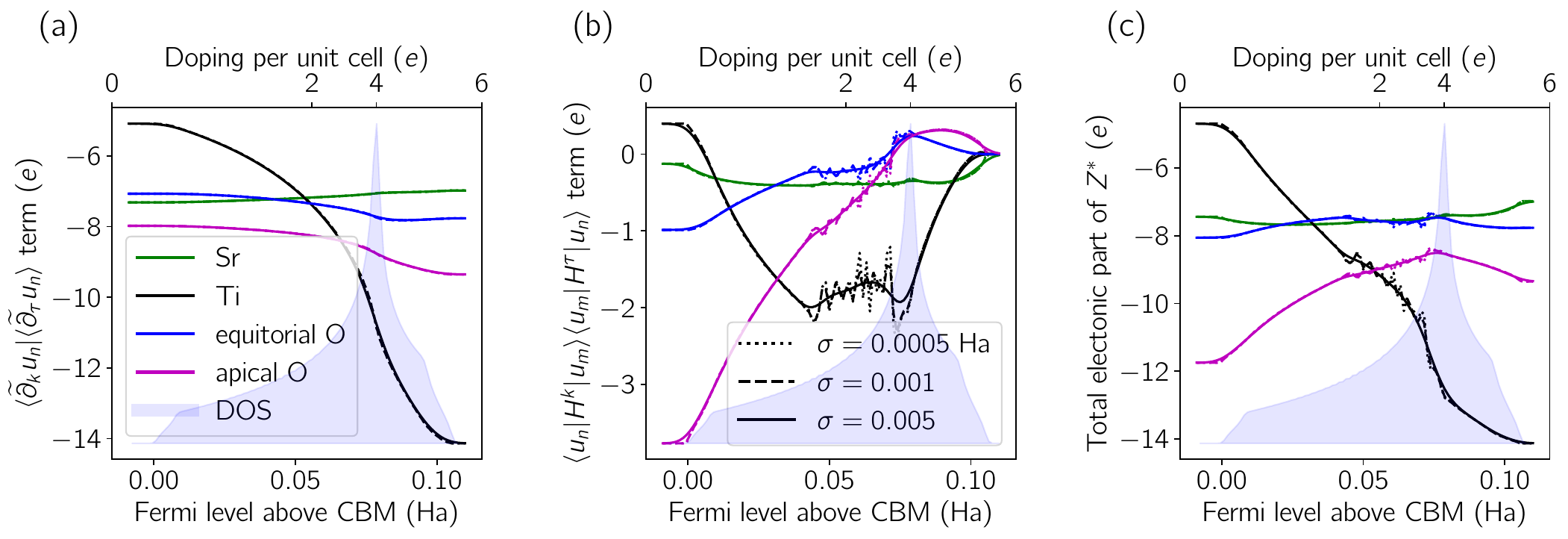}
\caption{\label{BEC_sigma} Same as Fig.~\ref{BEC_STO_kpts}, demonstrating convergence with respect to thermal spearing,  $\sigma$ ($k$ mesh is fixed to $100\times100\times100$).  }
\end{figure}

In this section we explore the numerical convergence of the naBECs and DW. In Fig.~\ref{BEC_STO_kpts} we plot the convergence of the electronic part of the naBECs with interpolated $k$ mesh. We plot separately the second [Fig.~\ref{BEC_STO_kpts}(a)] and third [Fig.~\ref{BEC_STO_kpts}(b)] terms in Eq.~(\ref{Zimp}) (the first ionic term is not included). We can see that the third term, which is nonzero only when empty bands are included in the active space, is the most difficult to converge especially as the van Hove singularity (vHS) is approached. In Fig.~\ref{BEC_sigma}, we plot the convergence of the naBEC for STO with respect to the width of the Gaussian smearing $\sigma$. Similarly to the case with $k$ mesh, the third term in Eq.~(\ref{Zimp}) shows oscillations for lower meshes as the vHS is approached, that are smoothed with larger smearings. These oscillations could also be removed with larger $k$ meshes.

We turn now to the convergence of Drude weight (DW), and the naBEC sum rule. In the main text, two versions of of the DW are discussed, i.e., whether the perturbation and response  are both taken to be the \emph{velocity} operators (``velocity-velocity'', which we will denote here as $D$), or the velocity operator and \emph{momentum} operator (``velocity-momentum'', which we will denote here as $\widetilde{D}$). The velocity and momentum operators differ in our density-functional theory calculations since the pseudopotentials contain nonlocal potentials \cite{Dreyer2018,Pickard2003,ICL2001, Lu2012}. 

\begin{figure}
\includegraphics[width=\columnwidth]{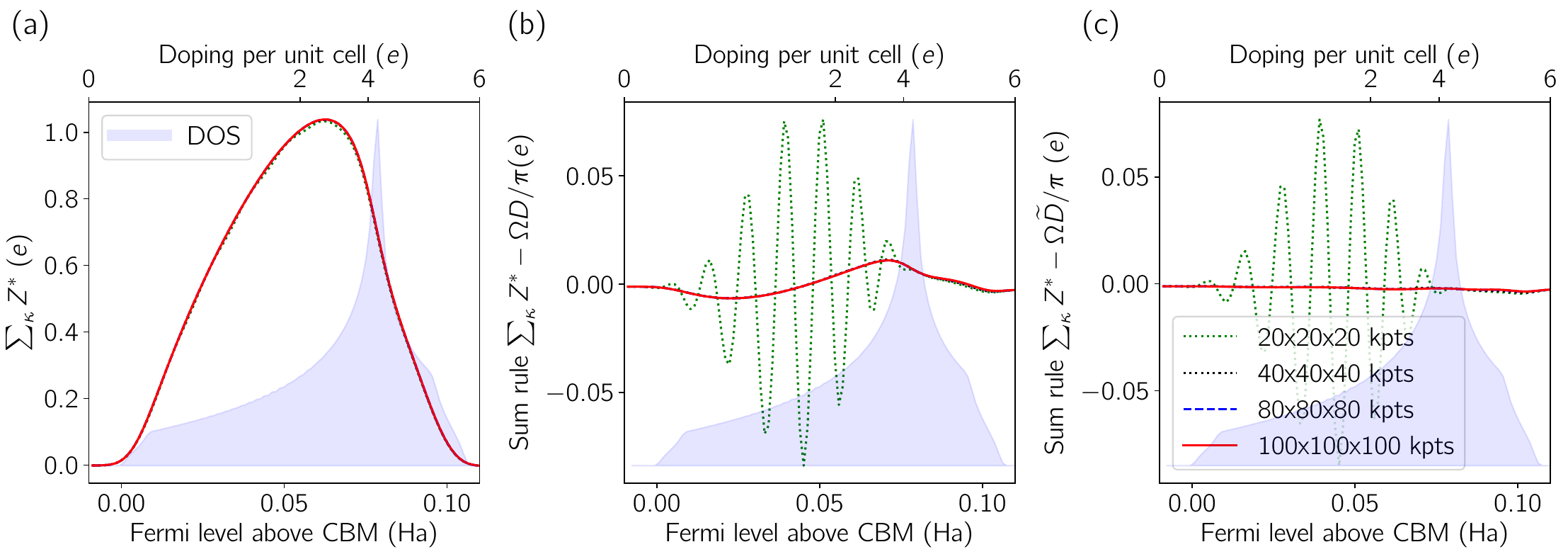}
\caption{\label{SumRule_STO_kpts}  (a) Sum of nonadiabatic Born effective charges (naBECs), (b) sum of naBECs and standard velocity-velocity Drude weight (DW), (c) sum of naBECs and modified momentum-velocity version of DW, for different $k$ meshes, with the density of states of the Ti $t2g$ manifold superimposed. We set the thermal smearing to $\sigma=0.005$ Ha }
\end{figure}

\begin{figure}
\includegraphics[width=\columnwidth]{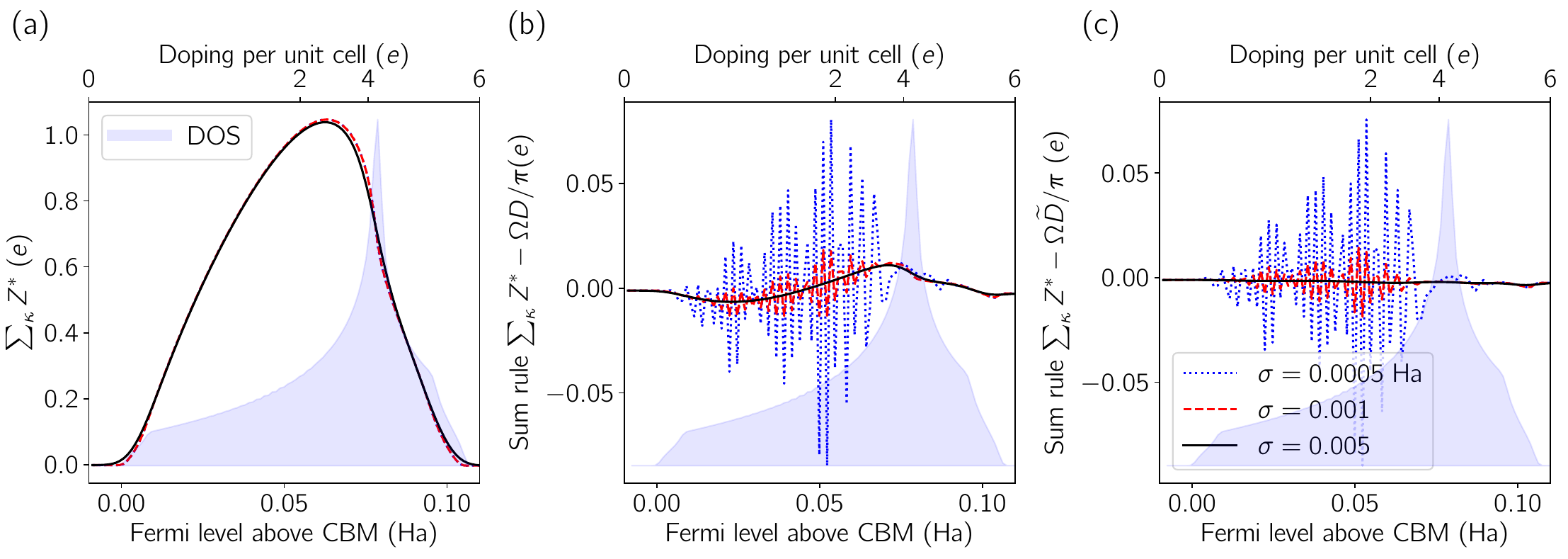}
\caption{\label{SumRule_STO_sigma} Same as Fig.~\ref{SumRule_STO_kpts}, but for different values of the thermal smearing $\sigma$ with the $k$-mesh set to $100\times100\times100$.  }
\end{figure}

In Fig.~\ref{SumRule_STO_kpts}(a) we plot the sublattice sum of the naBEC (including the ionic contribution) for different interpolated $k$ meshes; we see that this sum converges quite rapidly. In Fig.~\ref{SumRule_STO_kpts}(b) and (c) we demonstrate the naBEC sum rule. This also converges quickly with $k$-mesh. Also, we clearly see that the standard DW violates the naBEC sum rule, which is accuratly satisified by the $\widetilde{D}$ version. In Fig.~\ref{SumRule_STO_sigma}, we plot the convergence of the same quantities with the smearing. As with the naBECs, we see that a $\sigma$ of 0.005 Ha is required to smooth the oscillations in the sum rule for a $k$ mesh of $100\times100\times100$. A larger mesh would allow us to use a smaller smearing.

\begin{figure}
\includegraphics[width=\columnwidth]{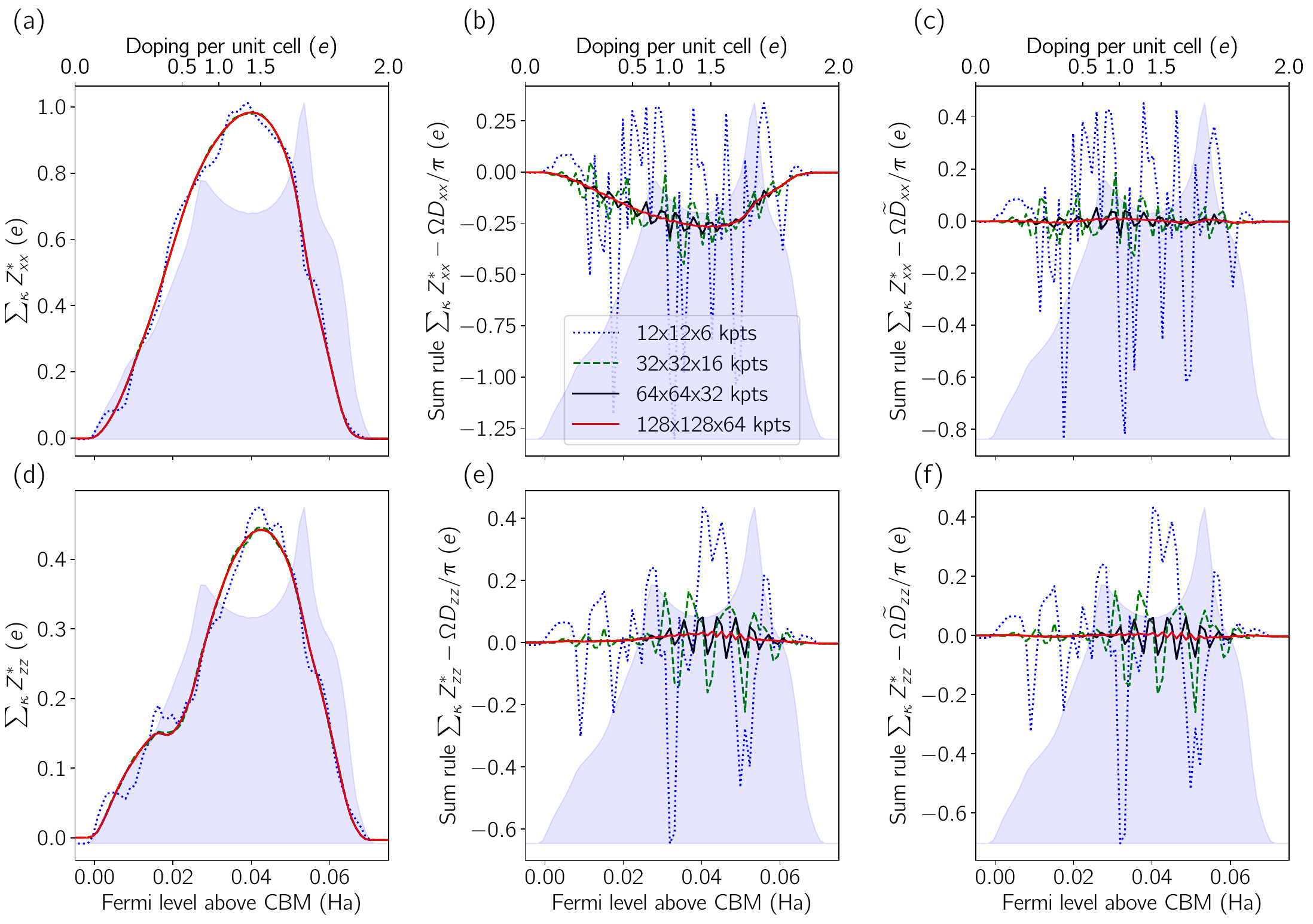}
\caption{\label{SumRule_sns_kpts} For SnS$_2$ using different $k$ meshes, sum of nonadiabatic Born effective charges (naBECs) for the (a) $xx$ and (d) $zz$ components; (b) and (e):  sum of naBECs and velocity-velocity Drude weight (DW); (c) and (f): sum of naBECs and momentum-velocity version of DW. The Density of states of the isolated conduction band superimposed. $\sigma=0.001$ Ha for all calculations.}
\end{figure}

The convergence behavior is overall quite similar for \sns{}. For example, in Fig.~\ref{SumRule_sns_kpts} we show the convergence of the naBEC sum rule with interpolated $k$ mesh for both the $\alpha=\beta=\hat{x}$ [Fig.~\ref{SumRule_sns_kpts}(a)-(c)] and $\alpha=\beta=\hat{z}$ [Fig.~\ref{SumRule_sns_kpts}(d)-(f)] components. We see the same behavior as for STO, where the naBEC sum rule is most accuratly satisfied for the $\widetilde{D}$ version of the DW.

\bibliography{flexo}